# Empowering deep neural quantum states through efficient optimization

Received: 20 March 2023

Accepted: 28 May 2024

Published online: 1 July 2024

Check for updates

Ao Chen & Markus Heyl

Computing the ground state of interacting quantum matter is a long-standing challenge, especially for complex two-dimensional systems. Recent developments have highlighted the potential of neural quantum states to solve the quantum many-body problem by encoding the many-body wavefunction into artificial neural networks. However, this method has faced the critical limitation that existing optimization algorithms are not suitable for training modern large-scale deep network architectures. Here, we introduce a minimum-step stochastic-reconfiguration optimization algorithm, which allows us to train deep neural quantum states with up to $10^6$ parameters. We demonstrate our method for paradigmatic frustrated spin-1/2 models on square and triangular lattices, for which our trained deep networks approach machine precision and yield improved variational energies compared to existing results. Equipped with our optimization algorithm, we find numerical evidence for gapless quantum-spin-liquid phases in the considered models, an open question to date. We present a method that captures the emergent complexity in quantum many-body problems through the expressive power of large-scale artificial neural networks.

It has been an ever-persisting quest in condensed-matter and quantum many-body physics to capture the essence of quantum many-body systems that is covered behind their exponential complexity. Although many numerical methods have been developed to access the quantum many-body problem with strong interactions, it still remains an extraordinary challenge to obtain accurate ground-state solutions, especially for complex and large two-dimensional systems. The respective challenges depend on the method utilized, such as the 'curse of dimensionality' in exact diagonalization[1], the notorious sign problem[2] in quantum Monte Carlo approaches[3] or the growth of entanglement and matrix contraction complexity in tensor network methods[4]. One of the paradigmatic instances of such complex two-dimensional quantum matter is the putative quantum-spin-liquid (QSL) phase in frustrated magnets[5]. Although a large variety of different numerical methods have been applied, the nature of many of the presumed QSLs still remains debated, such as the prototypical frustrated Heisenberg $J_1$–$J_2$ magnets on square[6–12] or triangular lattices[13–22].

Recently, neural quantum states (NQSs) have been introduced as a promising alternative for solving the quantum many-body problem by means of artificial neural networks[23]. This approach has already seen tremendous progress for QSLs[24–26]. However, this method also faces an outstanding challenge critically limiting its capabilities and its potential to date. Due to the rugged quantum landscape[27] with many saddle points, it is typically necessary to utilize stochastic reconfiguration (SR)[28] in the optimization. SR is a quantum generalization of natural gradient descent[29] and has a $\mathcal{O}(N_p^3)$ complexity for a network with $N_p$ parameters, which impedes the training of deep networks. Consequently, the current applications of NQS mainly focus on shallow networks, such as a restricted Boltzmann machine (RBM)[23,30] or shallow convolutional neural networks (CNNs)[25,31] with no more than ten layers and around $10^3$ parameters. Many efforts have been made to overcome the optimization difficulty in deep NQS based on either iterative solvers[23], approximate optimizers[32–36] or large-scale supercomputers[37,38]. However, the cost of SR still represents the key

Center for Electronic Correlations and Magnetism, University of Augsburg, Augsburg, Germany. e-mail: chenao.phys@gmail.com; markus.heyl@uni-a.de

limitation in increasing the network size and, thereby, fully materializing the exceptional power of artificial neural networks for outstanding physics problems.

In this work, we introduce an alternative training algorithm for NQS, which we term the minimum-step stochastic reconfiguration (MinSR). We show that the optimization cost in MinSR is reduced massively while it remains as accurate as SR. Concretely, the training cost of MinSR is only linear in $N_p$, which represents an enormous acceleration compared to SR. This, in turn, allows us to push the NQS towards the deep era by training deep networks with up to 64 layers and $10^6$ parameters. We apply our resulting algorithm to paradigmatic two-dimensional quantum spin systems, such as the spin-1/2 Heisenberg $J_1$–$J_2$ model, both to demonstrate the resulting accuracies for large system sizes beyond what is achievable with other computational methods and to address an outstanding question relating to the gaps in the model's QSL phases.

# Results

## Minimum-step stochastic reconfiguration

In the NQS approach, a neural network is utilized to encode and compress the many-body wavefunction. In a system with $N$ spin-1/2 degrees of freedom, the Hilbert space can be spanned by the $S_z$ spin configuration basis $|\sigma\rangle = |\sigma_1, \ldots, \sigma_N\rangle$ with $\sigma_i = \uparrow$ or $\downarrow$. An NQS with parameters $\theta$ maps every $\sigma$ at the input to a wavefunction component $\psi_{\theta,\sigma}$ at the output[23], as shown in Fig. 1a. The full quantum state is then given by the superposition $|\Psi_\theta\rangle = \sum_\sigma \psi_{\theta,\sigma} |\sigma\rangle$. When searching for ground states based on a variational Monte Carlo method (VMC), $\theta$ is optimized to minimize the variational energy $E_\theta = \langle\Psi_\theta| \mathcal{H} |\Psi_\theta\rangle / \langle\Psi_\theta|\Psi_\theta\rangle$.

The standard numerical approach for finding the minimal variational energy for NQS is SR. This is done by approximately implementing imaginary-time evolution. Thus, as the training progresses, the contributions from eigenstates with higher energies are systematically reduced, thereby pushing the state towards the ground state step by step. In every training step, this requires minimizing the quantum distance $d$ between the new variational state $|\Psi_{\theta+\delta\theta}\rangle$ and the exact imaginary-time evolved state $e^{-\mathcal{H}\delta\tau} |\Psi_\theta\rangle$, where $\delta\tau$ is the imaginary-time interval.

As proven in the Supplementary Information, the quantum distance $d$ can be estimated for a group of samples $\sigma$ with $P_\sigma \propto |\psi_\sigma|^2$ as $d^2 = \sum_\sigma \left|\sum_k \overline{O}_{\sigma k}\delta\theta_k - \bar{\epsilon}_\sigma\right|^2$, where $\sum_\sigma$ is performed on spin configurations in samples. We adopt the following notation: $\overline{O}_{\sigma k} = (O_{\sigma k} - \langle O_{\sigma k}\rangle)/\sqrt{N_s}$ with $O_{\sigma k} = \frac{1}{\psi_\sigma}\frac{\partial \psi_\sigma}{\partial \theta_k}$, and $\bar{\epsilon}_\sigma = -\delta\tau\left(E_{\mathrm{loc},\sigma} - \langle E_{\mathrm{loc},\sigma}\rangle\right)/\sqrt{N_s}$ with local energy $E_{\mathrm{loc},\sigma} = \sum_{\sigma'} \frac{\psi_{\sigma'}}{\psi_\sigma} H_{\sigma\sigma'}$, where $N_s$ is the number of samples and $\langle\ldots\rangle$ represents the mean value over the given set of samples.

Thus, the quantum distance $d$ can be rewritten as $d = ||\overline{O}\delta\theta - \bar{\epsilon}||$ if we treat $\delta\theta$ and $\bar{\epsilon}$ as vectors and $\overline{O}$ as a matrix. As a key consequence, we introduce a new linear equation

$$\overline{O}\delta\theta = \bar{\epsilon}, \tag{1}$$

whose least-squares solution minimizes the quantum distance $d$ and leads to the SR equation. Conceptually, one can understand the left-hand side of this equation as the change of the variational state induced by an optimization step of the parameters, and the right-hand side as the change of the exact imaginary-time evolving state. The traditional SR solution minimizing their difference is

$$\delta\theta = S^{-1}\overline{O}^\dagger\bar{\epsilon} \quad \text{with } S = \overline{O}^\dagger\overline{O}. \tag{2}$$

As illustrated in Fig. 1a, the matrix $S$ in equation (2) plays an important role as the quantum metric in VMC[29,39,40], which links the variations in the Hilbert space and the parameter space. However, inverting the matrix $S$, which has $N_p \times N_p$ elements, has $\mathcal{O}(N_p^3)$ complexity, and this a major difficulty when optimizing deep NQSs with large $N_p$. To reduce the cost of SR, we focus on a specific optimization case of a deep network with a large number of parameters $N_p$ but a relatively small amount of batch samples $N_s$, as occurs in most deep learning research. In this case, as shown in Fig. 1b, the rank of the $N_p \times N_p$ matrix $S$ is at most $N_s$, meaning that $S$ contains much less information than its capacity. As a more efficient way to express the information of the quantum metric, we introduce the neural tangent kernel $T = \overline{O}\,\overline{O}^\dagger$ (ref. 41), which has the same non-zero eigenvalues as $S$ but the matrix size reduces from $N_p \times N_p$ to $N_s \times N_s$.

As derived in Methods, we propose a new method termed MinSR using $T$ as the compressed matrix,

$$\delta\theta = \overline{O}^\dagger T^{-1}\bar{\epsilon} \quad \text{with } T = \overline{O}\,\overline{O}^\dagger, \tag{3}$$

which is mathematically equivalent to the traditional SR solution but only has $\mathcal{O}(N_p N_s^2 + N_s^3)$ complexity. For large $N_p$, it provides a tremendous acceleration with a time cost proportional to $N_p$ instead of $N_p^3$. Therefore, it can be viewed as a natural reformulation of traditional SR, which is particularly useful in the limit $N_p \gg N_s$, as relevant in deep learning situations. For a performance comparison, Extended Data Fig. 1 shows the time cost and accuracy of different optimization methods.

## Benchmark models

To demonstrate the exceptional performance of MinSR, we consider in the following the paradigmatic spin-1/2 Heisenberg $J_1$-$J_2$ model on a square lattice. This choice serves two purposes. On the one hand, this model serves as a standard benchmark system in various NQS studies and provides a convenient comparison to other state-of-the-art methods. On the other hand, it represents a paradigmatic reference case of QSLs in frustrated magnets, as an outstanding question regarding the nature of the QSL phase is whether it is gapped or gapless. The Hamiltonian of the system is given by

$$\mathcal{H} = J_1 \sum_{\langle i,j\rangle} \mathbf{S}_i\cdot\mathbf{S}_j + J_2 \sum_{\langle\langle i,j\rangle\rangle} \mathbf{S}_i\cdot\mathbf{S}_j, \tag{4}$$

where $\mathbf{S}_i = (S_i^x, S_i^y, S_i^z)$ with $S_i^x, S_i^y, S_i^z$ spin-1/2 operators at site $i$, $\langle i,j\rangle$ and $\langle\langle i,j\rangle\rangle$ indicate pairs of nearest-neighbour and next-nearest-neighbour sites, respectively, and $J_1$ is chosen to be equal to 1 for simplicity in this work.

We will specifically focus on two points in the parameter space: $J_2/J_1 = 0$ and $J_2/J_1 = 1/2$. At $J_2/J_1 = 0$, the Hamiltonian reduces to the non-frustrated Heisenberg model. At $J_2/J_1 = 1/2$, the $J_1$-$J_2$ model becomes strongly frustrated close to the maximally frustrated point where the system resides in a QSL phase[24], which imposes a great challenge for existing numerical methods, including NQS[31,42]. Two different designs of residual neural networks (ResNet), whose details we describe in Methods, will be employed for variationally learning the ground states of these benchmark models. A direct comparison with exact diagonalization results for the 6 × 6 square lattice can be found in Extended Data Fig. 2, which shows that our network can even approach machine precision on modern GPU and TPU hardware.

For a non-frustrated Heisenberg model of a 10 × 10 square lattice, a deep NQS trained by MinSR provides an unprecedentedly precise result that is better than all existing variational methods, as shown in Fig. 2a. The adopted reference ground-state energy per site is $E_{GS}/N = -0.67155267(5)$, as given by a simulation based on a stochastic series expansion[43] performed by ourselves, instead of the commonly used reference $E/N = -0.671549(4)$ from ref. 44 because our best NQS variational energy $E/N = -0.67155260(3)$ provides even better accuracy compared to this common reference energy. Thanks to the deep network architecture and the efficient MinSR method, the relative error of the variational energy $\epsilon_{\mathrm{rel}} = (E - E_{GS})/|E_{GS}|$ drops much faster than for the one-layer RBM as $N_p$ increases and finally reaches a level of $10^{-7}$, greatly outperforming existing results.

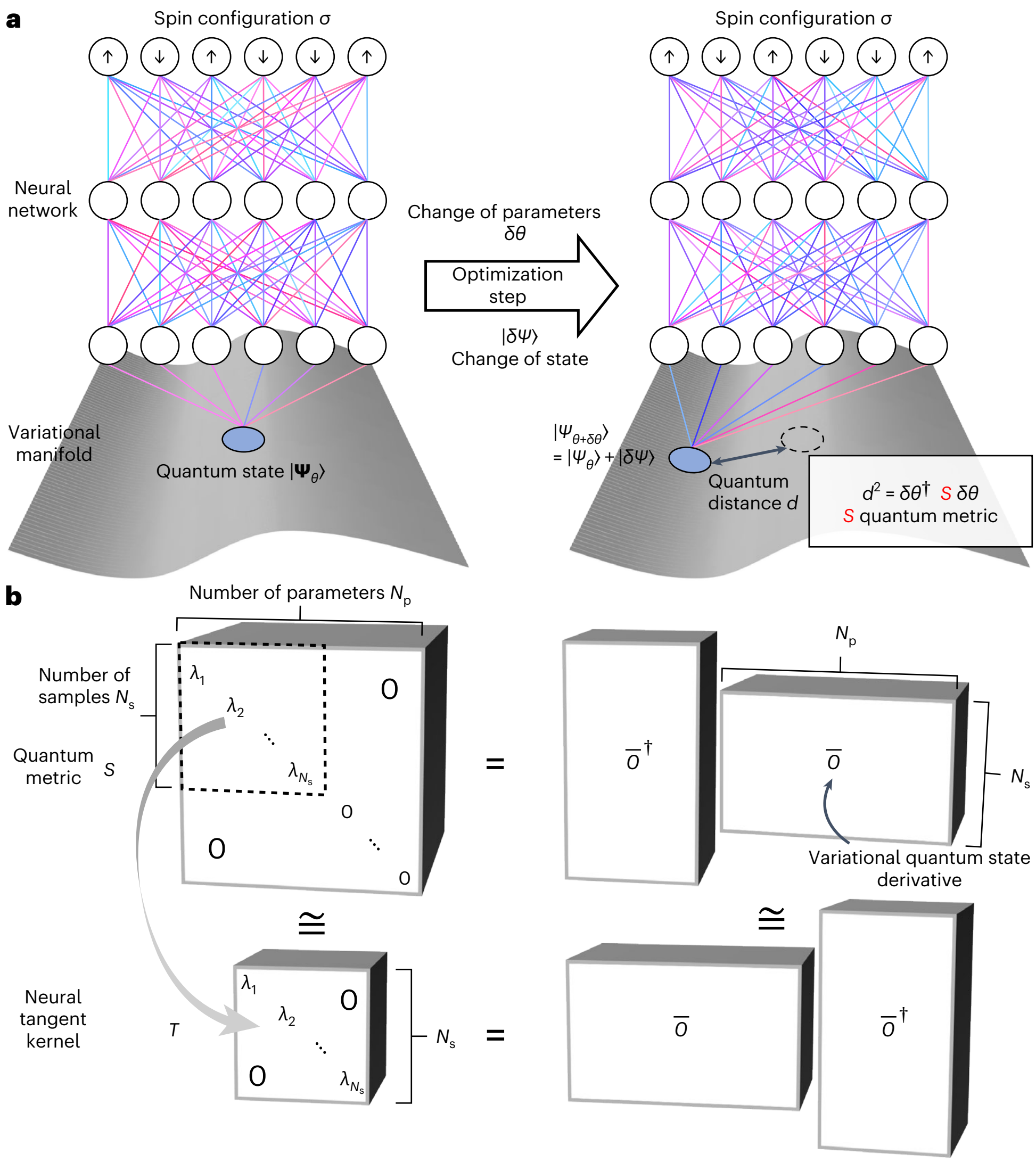

**Fig. 1 | Illustration of NQS and MinSR. a**, In the NQS approach, an artificial neural network is used to represent a quantum many-body state. A change of the network parameters for the NQS leads to a new quantum state, whose distance to the previous NQS is given by the quantum metric $S \in \mathbb{C}^{N_p \times N_p}$, where $N_p$ is the number of variational parameters. **b**, The quantum metric $S = \overline{O}^\dagger \overline{O}$ can be decomposed into a smaller matrix $\overline{O} \in \mathbb{C}^{N_s \times N_p}$ with $N_s \ll N_p$ the number of Monte Carlo samples. The optimization of an NQS involves the inversion of the quantum metric $S$, which is equivalent to determining its non-zero eigenvalues $\lambda_i$ with $i = 1, \ldots, N_s$. In MinSR, a neural tangent kernel $T = \overline{O}\,\overline{O}^\dagger \in \mathbb{C}^{N_s \times N_s}$ is introduced with identical eigenvalues $\lambda_i$ and, therefore, the essential information of $S$.

To attain the next level of complexity, we will now focus on the frustrated $J_1$–$J_2$ model, whose accurate ground-state solution has remained a key challenge for all available computational approaches. Figure 2b shows that, for a 10 × 10 square lattice, our method based on MinSR allows us to reach ground-state energies below what is possible with any other numerical scheme so far. In this context, the Marshall sign rule (MSR) limit shows the energy one can obtain without considering any frustration. As shown in the figure, the use of deep NQS becomes absolutely crucial as the shallow CNN is not guaranteed to beat the MSR limit. Most importantly, the variational energy we obtained was reduced upon increasing the network size for both networks trained by MinSR. We finally trained unprecedentedly large networks with 64 convolutional layers in ResNet1 and more than one million parameters in ResNet2, to attain the best variational energy $E/N = -0.4976921(4)$, which outperforms all existing numerical results. The extraordinary variational outcomes allow us to accurately estimate the ground-state energy $E_{GS}/N = -0.497715(9)$ by zero-variance extrapolation, as described in Methods. Compared with the previous best result[24], $\epsilon_{rel}$ in our biggest network is around 4 times lower, suggesting that our deep NQS result is substantially more accurate. From this, we conclude that the deep NQS trained by MinSR is superior even in the frustrated case, which was argued to be challenging for NQS on a general level[45]. The variational energies of different methods in this prototypical model are summarized in Extended Data Table 1.

Finally, we aim to provide evidence that our approach still exhibits advantageous performance compared to other computational methods upon further increasing the system size. Figure 2c presents the variational energy obtained for a 16 × 16 square lattice and compares the results with existing results in the literature. One can clearly see that our approach yields the best variational energy $E/N = -0.4967163(8)$ for the frustrated $J_1$-$J_2$ model on such a large lattice. Compared with the best existing variational result given in ref. 37, $\epsilon_{rel}$ in this work is still $2.5 \times 10^{-4}$ lower. In summary, the deep NQS trained by MinSR provides results for large frustrated models that are not only on a par with other state-of-the-art methods but can substantially outperform them.

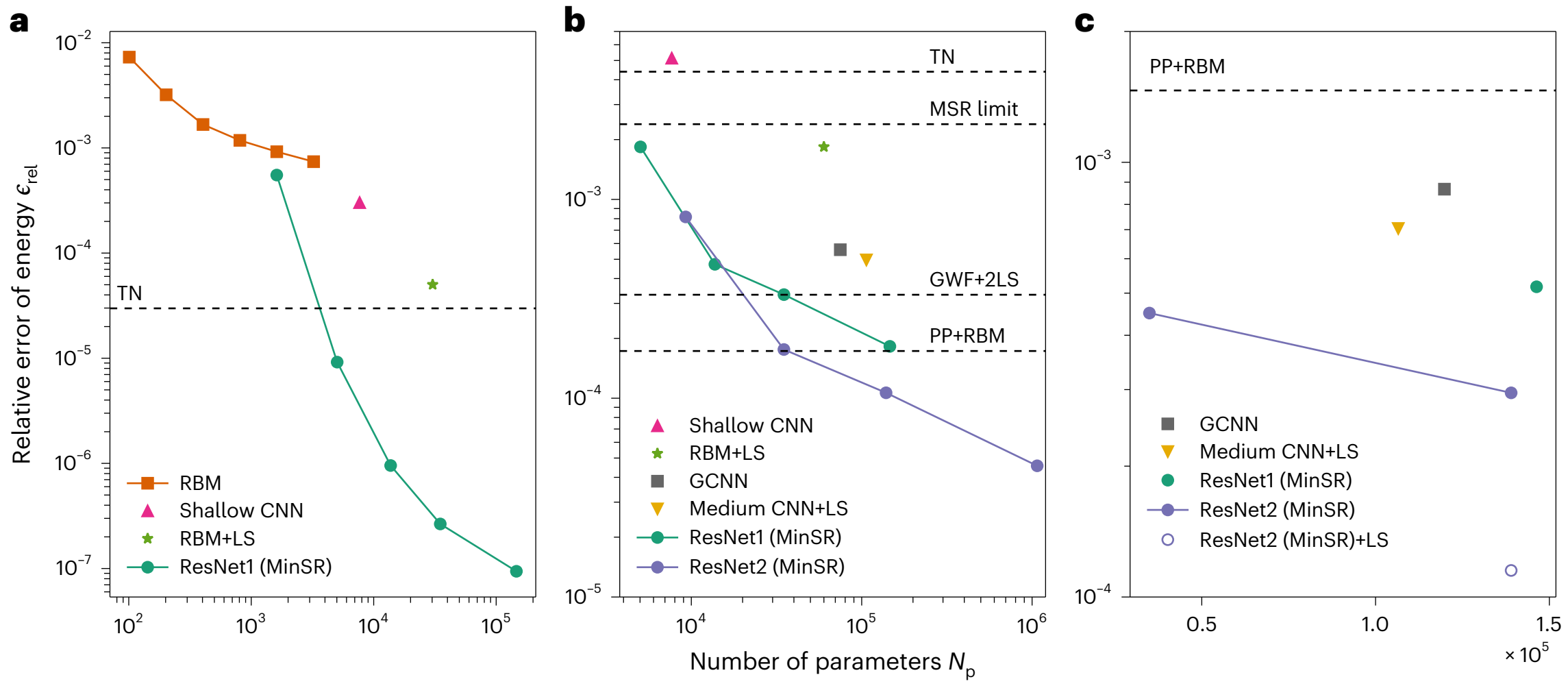

**Fig. 2 | Relative error of the variational energy $\epsilon_{rel} = (E - E_{GS})/|E_{GS}|$ for a square lattice, where $E_{GS}$ is the exact ground-state energy estimated by stochastic series expansion in the non-frustrated case and zero-variance extrapolation in the frustrated case. a**, Non-frustrated 10 × 10 Heisenberg model. The variational energies obtained in this work by using a deep ResNet trained with MinSR are compared to previous results in the literature including an RBM[23], shallow CNN[31] and RBM with a Lanczos step (RBM+LS)[38]. As no tensor network (TN) data are available for the periodic boundary condition, the best result with an open boundary condition is included as a dashed line[51]. **b**, Frustrated 10 × 10 $J_1$-$J_2$ model at $J_2/J_1 = 0.5$. The results obtained in this work with MinSR for two designs of ResNet are compared to previous results in the literature for a shallow CNN[31], RBM+LS[38], group convolutional neural network (GCNN)[26] and medium CNN[37]. Further results from methods other than NQS are included as dashed lines, such as a tensor network[9], the Gutzwiller wavefunction with two Lanczos steps (GWF+2LS)[8], and a combination of the pair product state and RBM (PP+RBM)[24]. As a further reference, the so-called MSR limit is included. This was obtained from an NQS trained for a wavefunction where the sign structure was not learned but rather fixed by the MSR. **c**, Frustrated 16 × 16 $J_1$–$J_2$ model at $J_2/J_1 = 0.5$.

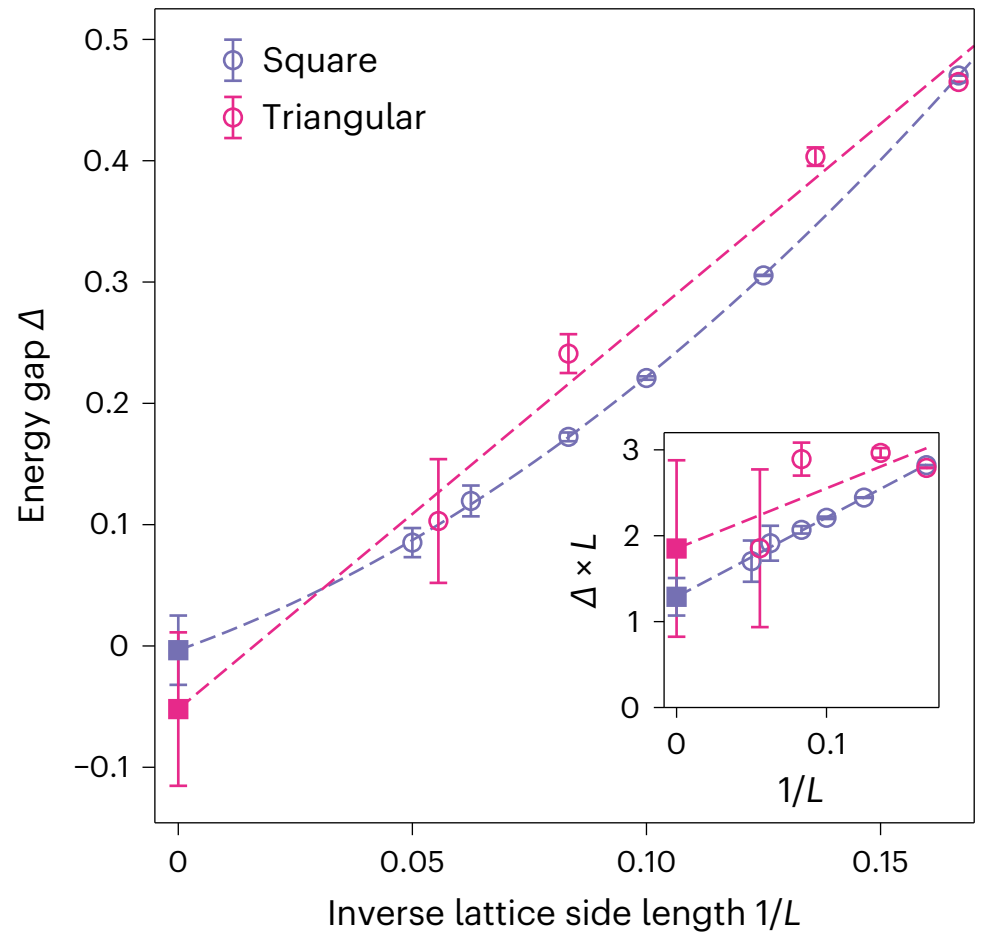

**Fig. 3 | Energy gap $\Delta$ between the ground state with total spin $S = 0$ and the excited state with $S = 1$ as a function of inverse linear length $1/L$ at the maximally frustrated point.** The inset includes the behaviour of the rescaled gap $\Delta \times L$ versus $1/L$.

## Energy gaps of a QSL

Although so far we have focused on demonstrating the exceptional performance of the MinSR method, we now take the next step by addressing an outstanding physical question regarding the $J_1$-$J_2$ Heisenberg model considered. Concretely, we utilize the combination of the deep NQS and MinSR to study the gaps for two famous QSL candidates in the $J_1$-$J_2$ model on a square lattice and on a triangular lattice. In these systems, several works in the literature[6–22] have shown the existence of QSL phases, although the energy gaps in the thermodynamic limit, especially for the triangular lattice, still remain debated. Figure 3 present an extrapolation of the energy gaps between states with total spin $S = 0$ and $S = 1$ to the thermodynamic limit within the most frustrated regime in which QSL candidates reside. As explained in Extended Data Figs. 3 and 4, the energy is estimated by NQS trained by MinSR with a Lanczos step and zero-variance extrapolation to increase accuracy. In the Supplementary Information, we provide the spin and dimer structure factors to support the existence of a QSL phase on the triangular lattice and compare gap estimates with and without zero-variance extrapolation.

On the square lattice, the gaps are measured in the total spin $S = 1$ sector and momentum $k = (\pi, \pi)$ ($M$-point) at the most frustrated point $J_2/J_1 = 0.5$ for different system sizes, including 6 × 6, 8 × 8, 10 × 10, 12 × 12, 16 × 16 and 20 × 20. As shown by the small fitting error in Fig. 3 with $\Delta = a + b/L + c/L^2$, the vanishing gap $\Delta = 0.00(3)$ in the thermodynamic limit provides an unprecedented precision and is so far the most accurate extrapolation at this most frustrated point. In addition to the direct extrapolation of the energy gap $\Delta$, we support our finding of a vanishing gap in the inset of Fig. 3, which display $\Delta \times L$ as a function of $1/L$ (ref. 24). Although a finite gap would imply a divergent $\Delta \times L$, we observe a constant value, further corroborating our conclusion of a gapless phase in the thermodynamic limit. Combined with the large lattice sizes used, this result shows strong evidence of gapless QSLs as suggested by refs. 10–12,24 in contrast to the conclusion of the gapped QSLs in ref. 6.

The triangular $J_1$–$J_2$ model has even stronger frustration compared to the square one, leading to larger variational errors in different methods and more controversy regarding the nature of the QSLs. To target the QSLs in this model, we also studied the most frustrated point at $J_2/J_1 = 0.125$. The gaps were measured for the $S = 1$ and $k = (4\pi/3, 0)$ state on lattices 6 × 6, 6 × 9, 12 × 12 and 18 × 18 for the triangular lattice. Due to the larger variational error on the triangular lattice compared to the square case, a linear fitting $\Delta = a + b/L$ was utilized instead of the quadratic one to prevent overfitting. For a lattice with unequal extents $L_x$ and $L_y$ in different dimensions, $L$ is defined as $\sqrt{L_x L_y}$. Our data matches well with the linear relation $\Delta \propto 1/L$ as expected for Dirac spin liquids, and the vanishing gap at the thermodynamic limit is $\Delta = -0.05(6)$. Furthermore, we also performed an extrapolation of $\Delta \times L$ (inset of Fig. 3). We found a finite $\Delta \times L$ upon increasing the system size $L$, indicating a vanishing gap in the thermodynamic limit. We take these results as strong numerical evidence suggesting the existence of a gapless QSL as also indicated in refs. 13,16,20,22 instead of a gapped QSL in refs. 14,15,21. Consequently, these numerical results demonstrate the exceptional computational power of the MinSR method applied to NQS wavefunctions, especially for the challenging regime of frustrated quantum magnets in two dimensions.

## Discussion

To date, there have been tremendous efforts in solving quantum many-body problems in two major directions, studying the simplified Hilbert space given by specific physical backgrounds on classical computers and traversing the full Hilbert space on quantum computers. In this work, we present another promising approach that is supported by deep NQSs. This method allows us to approximate the complexity of quantum many-body problems through the emergent expressive power of large-scale neural networks.

For the future, we envision promising research directions, for instance, studying fermionic systems including the celebrated Hubbard model[46,47] or ab initio quantum chemistry[48], in which the traditional methods have limited accuracy, especially in the strongly interacting regime. Moreover, it is key to point out that the MinSR method is not at all restricted to NQS. As a general optimization method in VMC, it can also be applied to other variational wavefunctions, like tensor networks, so that a more complex ansatz can be introduced in these conventional methods to enhance the expressivity. It will also be of great importance to exploit the expressive power of large-scale variational wavefunctions through a suitable design that would lower the computational cost and increase the accuracy.

We can further envision the application of MinSR beyond the scope of physics for general machine learning tasks, if a suitable space for optimization like the Hilbert space in physics can be defined for which we can construct an equation like equation (1). In reinforcement learning tasks, for instance, obtaining gradients from the action in the environment is usually the most time-consuming part of the training, so a MinSR-like natural policy gradient[49] can provide more accurate optimization directions without substantially increased time cost and greatly improve the training efficiency, even for very deep neural networks. Recently, a method inspired by MinSR has already found applications in general machine learning tasks[50].

## Online content

Any methods, additional references, Nature Portfolio reporting summaries, source data, extended data, supplementary information, acknowledgements, peer review information; details of author contributions and competing interests; and statements of data and code availability are available at https://doi.org/10.1038/s41567-024-02566-1.

# Methods

## Derivation of the MinSR equation

MinSR was derived based on the observation that equation (1) is underdetermined when $N_s < N_p$. To obtain a unique $\delta\theta$ solution, we employed the least-squares minimum-norm condition, which is widely used for underdetermined linear equations. To be specific, we chose, among all solutions with minimum residual error $||\overline{O}\delta\theta - \bar{\epsilon}||$, the one minimizing the norm of the variational step $||\delta\theta||$, which helps to reduce higher-order effects, prevent overfitting and improve stability. We called this method MinSR due to the additional minimum-step condition. In this section, we adopt two different approaches, namely the Lagrangian multiplier method and the pseudo-inverse method, to derive the MinSR formula in equation (3).

**Lagrangian multiplier.** The MinSR solution can be derived by minimizing the variational step $\sum_k |\delta\theta_k|^2$ under the constraint of minimum residual error $\sum_\sigma |\sum_k \overline{O}_{\sigma k}\delta\theta_k - \bar{\epsilon}_\sigma|^2$. To begin, we assume that the minimum residual error is 0, which can always be achieved by letting $N_s < N_p$ and assuming a typical situation in VMC that $\overline{O}_{\sigma k}$ values obtained by different samples are linearly independent. This leads to constraints $\sum_k \overline{O}_{\sigma k}\delta\theta_k - \bar{\epsilon}_\sigma = 0$ for each $\sigma$. The Lagrange function is then given by

$$\mathcal{L}(\{\delta\theta_k\},\{\alpha_\sigma\}) = \sum_k |\delta\theta_k|^2 - \left[\sum_\sigma \alpha_\sigma^* \sum_k (\overline{O}_{\sigma k}\delta\theta_k - \bar{\epsilon}_\sigma) + \text{h.c.}\right], \quad (5)$$

where $\alpha_\sigma$ is the Lagrangian multiplier. Written in matrix form, the Lagrangian function becomes

$$\mathcal{L}(\delta\theta,\alpha) = \delta\theta^\dagger\delta\theta - \alpha^\dagger(\overline{O}\delta\theta - \bar{\epsilon}) - (\delta\theta^\dagger \overline{O}^\dagger - \bar{\epsilon}^\dagger)\alpha. \quad (6)$$

From $\partial\mathcal{L}/\partial(\delta\theta^\dagger) = 0$, one obtains

$$\delta\theta = \overline{O}^\dagger \alpha. \quad (7)$$

Putting equation (7) back into $\overline{O}\delta\theta = \bar{\epsilon}$, one can solve $\alpha$ as

$$\alpha = (\overline{O}\,\overline{O}^\dagger)^{-1}\bar{\epsilon}. \quad (8)$$

Combining equation (8) with equation (7), one obtains the final solution as

$$\delta\theta = \overline{O}^\dagger(\overline{O}\,\overline{O}^\dagger)^{-1}\bar{\epsilon}, \quad (9)$$

which is the MinSR formula in equation (3). A similar derivation also applies when $\overline{O}$, $\delta\theta$ and $\bar{\epsilon}$ are all real.

In our simulations, the residual error is non-zero, which differs from our previous assumption. This is because the inverse in equation (9) is replaced by a pseudo-inverse with finite truncation to stabilize the solution in the numerical experiments.

**Pseudo-inverse.** To simplify the notation, we use $A = \overline{O}, x = \delta\theta$ and $b = \bar{\epsilon}$. We will prove that for a linear equation $Ax = b$,

$$x = A^{-1}b = (A^\dagger A)^{-1}A^\dagger b = A^\dagger(AA^\dagger)^{-1}b \quad (10)$$

is the least-squares minimum-norm solution, where the matrix inverse is pseudo-inverse.

First, we prove $x = A^{-1}b$ is the solution we need. The singular value decomposition of $A$ gives

$$A = U\Sigma V^\dagger, \quad (11)$$

where $U$ and $V$ are unitary matrices, and $\Sigma$ is a diagonal matrix with $\sigma_i = \Sigma_{ii} = 0$ if and only if $i > r$ with $r$ the rank of $A$. The least-squares solution is given by minimizing

$$\begin{aligned}||Ax - b||^2 &= ||U\Sigma V^\dagger x - b||^2 \\ &= ||\Sigma x' - b'||^2 \\ &= \sum_{i=1}^{r}\left(\sigma_i x_i' - b_i'\right)^2 + \sum_{i=r+1}^{N_s} b_i'^2,\end{aligned} \quad (12)$$

where $x' = V^\dagger x$, $b' = U^\dagger b$ and $N_s$ is the dimension of $b$, and the second step is because applying a unitary matrix does not change the norm of a vector. Therefore, all the least-squares solutions take the form

$$x_i' = \begin{cases} b_i'/\sigma_i, & i \le r, \\ \text{any value}, & i > r. \end{cases} \quad (13)$$

Among all these possible solutions, the one that minimizes $||x|| = ||x'||$ is

$$x_i' = \begin{cases} b_i'/\sigma_i, & i \le r, \\ 0, & i > r. \end{cases} \quad (14)$$

With the following definition of a pseudo-inverse

$$\begin{aligned}A^{-1} &= V\Sigma^+U^\dagger, \\ \Sigma^+_{ij} &= \delta_{ij} \times \begin{cases} 1/\sigma_i, & \sigma_i > 0, \\ 0, & \sigma_i = 0, \end{cases}\end{aligned} \quad (15)$$

we have $x' = \Sigma^+ b'$, so the final solution is

$$x = Vx' = V\Sigma^+U^\dagger b = A^{-1}b. \quad (16)$$

Furthermore, we show the following equality

$$A^{-1} = (A^\dagger A)^{-1}A^\dagger = A^\dagger(AA^\dagger)^{-1}. \quad (17)$$

With the singular value decomposition of $A$ in equation (11), equation (17) can be directly proved by

$$\begin{aligned}(A^\dagger A)^{-1}A^\dagger &= (V\Sigma U^\dagger U\Sigma V^\dagger)^{-1}V\Sigma U^\dagger \\ &= V(\Sigma^+)^2V^\dagger V\Sigma U^\dagger \\ &= V\Sigma^+U^\dagger \\ &= A^{-1},\end{aligned} \quad (18)$$

and

$$\begin{aligned}A^\dagger(AA^\dagger)^{-1} &= V\Sigma U^\dagger(U\Sigma V^\dagger V\Sigma U^\dagger)^{-1} \\ &= V\Sigma U^\dagger U(\Sigma^+)^2U^\dagger \\ &= V\Sigma^+U^\dagger \\ &= A^{-1}.\end{aligned} \quad (19)$$

In the derivation, the shapes of diagonal matrices $\Sigma$ and $\Sigma^+$ are not fixed but assumed to match their neighbour matrices to make the matrix multiplication valid.

Equation (17) shows that the SR solution in equation (2) and MinSR solution in equation (3) are both equivalent to the pseudo-inverse solution $\delta\theta = \overline{O}^{-1}\bar{\epsilon}$, which justifies MinSR as a natural alternative to SR when $N_s < N_p$.

## MinSR solution

**Numerical solution.** In this section, we focus on how to solve the MinSR equation numerically:

$$\delta\theta = \overline{O}^{\dagger} T^{-1}\bar{\epsilon}. \tag{20}$$

The whole computation, starting from $T = \overline{O}\,\overline{O}^{\dagger}$, should be executed under double-precision arithmetic to ensure that small eigenvalues are reliable.

Then a suitable pseudo-inverse should be applied to obtain a stable solution. In practice, the Hermitian matrix $T$ is first diagonalized as $T = UDU^{\dagger}$, and the pseudo-inverse is given by

$$T^{-1} = UD^{+}U^{\dagger}, \tag{21}$$

where $D^{+}$ is the pseudo-inverse of the diagonal matrix $D$, numerically given by a cutoff below which the eigenvalues are regarded as 0, that is

$$\lambda_i^{+} = \begin{cases} 1/\lambda_i, & |\lambda_i| \geq r_{\rm pinv}|\lambda_{\rm max}| + a_{\rm pinv}, \\ 0, & |\lambda_i| < r_{\rm pinv}|\lambda_{\rm max}| + a_{\rm pinv}, \end{cases} \tag{22}$$

where $\lambda_i$ and $\lambda_i^{+}$ are the diagonal elements of $D$ and $D^{+}$, $\lambda_{\rm max}$ is the largest value among $\lambda_i$, and $r_{\rm pinv}$ and $a_{\rm pinv}$ are the relative and absolute pseudo-inverse cutoffs. In most cases, we choose $r_{\rm pinv} = 10^{-12}$ and $a_{\rm pinv} = 0$. Furthermore, we modify the aforementioned direct cutoff to a soft one[52]:

$$\lambda_i^{+} = \left[\lambda_i\left(1 + \left(\frac{r_{\rm pinv}|\lambda_{\rm max}| + a_{\rm pinv}}{|\lambda_i|}\right)^6\right)\right]^{-1} \tag{23}$$

to avoid abrupt changes when the eigenvalues cross the cutoff during optimization.

**Complex neural networks.** Our original MinSR formula equation (3) can be applied when the network is real or complex holomorphic. In our ResNet2 architecture, however, the neural network parameters are real but the network outputs can be complex, in which case equation (3) cannot be directly applied. For other non-holomorphic networks, a complex parameter can be taken as two independent real parameters but this problem still occurs. To obtain the MinSR equation in these special cases, notice that the quantum distance $d$ between $|\Psi_{\theta+\delta\theta}\rangle$ and $e^{i\mathcal{H}\delta\tau}|\Psi_\theta\rangle$ can be reformulated as

$$\begin{aligned} d^2 &= ||\overline{O}\delta\theta - \bar{\epsilon}||^2 \\ &= ||\,\mathrm{Re}(\overline{O})\delta\theta - \mathrm{Re}(\,\bar{\epsilon}\,)||^2 + ||\,\mathrm{Im}(\overline{O})\delta\theta - \mathrm{Im}(\,\bar{\epsilon}\,)||^2, \end{aligned} \tag{24}$$

assuming $\overline{O}$ and $\bar{\epsilon}$ are complex while $\delta\theta$ is real. By defining

$$\overline{O}' = \begin{pmatrix} \mathrm{Re}\,\overline{O} \\ \mathrm{Im}\,\overline{O} \end{pmatrix}, \quad \bar{\epsilon}' = \begin{pmatrix} \mathrm{Re}\,\bar{\epsilon} \\ \mathrm{Im}\,\bar{\epsilon} \end{pmatrix}, \tag{25}$$

one can rewrite the quantum distance again as $d^2 = ||\overline{O}'\delta\theta - \bar{\epsilon}'||^2$ with all entities real. The MinSR solution, in this case, is similarly given by

$$\delta\theta = \overline{O}'^{\dagger} T'^{-1}\bar{\epsilon} \quad \text{with } T' = \overline{O}'\,\overline{O}'^{\dagger}. \tag{26}$$

Similar arguments can also provide the SR equation in the non-holomorphic case as

$$\begin{aligned} &\delta\theta = S'^{-1}F' \\ &\text{with } S' = \overline{O}'^{\dagger}\overline{O}' = \mathrm{Re}\,S,\ F' = \overline{O}'^{\dagger}\bar{\epsilon}' = \mathrm{Re}\,F, \end{aligned} \tag{27}$$

where $S = \overline{O}^{\dagger}\overline{O}$ and $F = \overline{O}^{\dagger}\bar{\epsilon}$ are the same as for the ordinary SR solution. This solution agrees with the widely used non-holomorphic SR solution[53].

## Neural quantum states

In this work, we adopt two different designs of ResNets. Several techniques are also applied to reduce the error.

**ResNet1.** The first architecture, as suggested in ref. 54, has two convolutional layers in each residual block, each given by a layer normalization, a ReLU activation function and a convolutional layer sequentially. All the convolutional layers are real-valued with the same number of channels and kernel size. After the forward pass through all residual blocks, a final activation function $f(x) = \cosh x\,(x > 0),\ 2 - \cosh x\,(x < 0)$ is applied, which resembles the $\cosh(x)$ activation in RBM but can also give negative outputs so that the whole network is able to express sign structures while still being real-valued. In the non-frustrated case, $|f(x)|$ is used as the final activation function to make all outputs positive. After the final activation function, the outputs $v_i$ are used to compute the wavefunction as $\psi_\sigma^{\rm net} = \prod_i (v_i/t)$, where $t$ is a rescaling factor updated in every training step. $t$ is used to prevent a data overflow after the product.

**ResNet2.** The second design of ResNet basically follows ref. 26. In this architecture, the residual blocks are the same as ResNet1 but the normalization layers are removed. In the last layer, two different kinds of activations can be applied. For real-valued wavefunctions, we chose $f(x) = \sinh(x) + 1$. For complex-valued wavefunctions, we split all channels in the last layer into two groups and employ $f(x_1, x_2) = \exp(x_1 + ix_2)$. A rescaling factor $t$ is also inserted in suitable places in $f$ to prevent an overflow.

Finally, a sum is performed to obtain the wavefunction. Considering the possible non-zero momentum $\mathbf{q}$, the wavefunction is given by

$$\psi_\sigma^{\rm net} = \sum_i e^{-i\mathbf{q}\cdot\mathbf{r}_i} \sum_c v_{c,i}, \tag{28}$$

where $v_{c,i}$ is the last-layer neuron at channel $c$ and site $i$, and $r_i$ is the real-space position of site $i$. This definition ensures that the whole NQS has a momentum $\mathbf{q}$.

In summary, ResNet1 performs better when one applies transfer learning from a small lattice to a larger one, but ResNet2, in general, has better accuracy and stability. Moreover, ResNet2 allows one to implement non-zero momentum, which is key to finding low-lying excited states.

**Sign structure.** On top of the raw output from the neural network $\psi_\sigma^{\rm net}$, the MSR[55] is applied to wavefunctions on a square lattice, which serves as the exact sign structure for the non-frustrated Heisenberg model but is still the approximate sign structure in the frustrated region around $J_2/J_1 \approx 0.5$. The sign structure representing the 120° magnetic order is also applied for the triangular lattice. Although these sign structures are additional physical inputs for specific models, the generality is not reduced because it has been shown that simple sign structures such as MSR can be exactly solved by an additional sign network[56,57].

**Symmetry.** Symmetry plays an important role in improving the accuracy and finding low-lying excited states for NQS[30,58]. In this work, we apply symmetry on top of the well-trained $\psi_\sigma^{\rm net}$ to project variational states onto suitable symmetry sectors. Assuming the system permits a symmetry group of order $|G|$ represented by operators $T_i$ with characters $\omega_i$, the symmetrized wavefunction is then defined as[30,59]

$$\psi_\sigma^{\rm symm} = \frac{1}{|G|}\sum_i \omega_i^{-1}\psi_{T_i\sigma}^{\rm net}. \tag{29}$$

With translation symmetry already enforced by the CNN architecture, the remaining symmetries applied by equation (29) are the point group symmetry, which is $C_{4v}$ for the square lattice and $D_6$ for the triangular lattice, and the spin inversion symmetry $\sigma \to -\sigma$ (refs. 60–64).

### Zero-variance extrapolation

The variational wavefunction provides an inexact estimate of the ground-state energy due to the variational error. Fortunately, in VMC one can compute the energy variance

$$\sigma^2 = \langle \mathcal{H}^2 \rangle - \langle \mathcal{H} \rangle^2 \tag{30}$$

as an estimate of the variational error. Hence, an extrapolation to zero energy variance gives a better estimate of the ground-state energy[65,66], which has been successfully applied to NQS in refs. 30,37. In the following, we adopt the derivation in ref. 66 to show how to perform the extrapolation.

Assuming the normalized variational state $|\psi\rangle$ deviates only slightly from the exact ground state $|\psi_g\rangle$, one can express it as

$$|\psi\rangle = \sqrt{1-\lambda^2}\,|\psi_g\rangle + \lambda\,|\psi_e\rangle, \tag{31}$$

where $|\psi_e\rangle$ represents the error in the variational state orthogonal to the ground state and $\lambda$ is a small positive number indicating the error strength. Denoting $E_g = \langle\psi_g|\mathcal{H}|\psi_g\rangle$, $E_e = \langle\psi_e|\mathcal{H}|\psi_e\rangle$ and $\langle\mathcal{H}^2\rangle_e = \langle\psi_e|\mathcal{H}^2|\psi_e\rangle$, one can express the variational energy as

$$E = \langle\psi|\mathcal{H}|\psi\rangle = E_g + \lambda^2(E_e - E_g). \tag{32}$$

Similarly, the energy variance can be written as

$$\sigma^2 = \lambda^2\left(\langle\mathcal{H}^2\rangle_e - 2E_gE_e + E_g^2\right) + \mathcal{O}(\lambda^4). \tag{33}$$

If the error state $|\psi_e\rangle$ does not change substantially in different training attempts, there is a linear relation

$$(E - E_g) \propto \sigma^2 \tag{34}$$

for small $\lambda$, so a linear extrapolation to $\sigma^2 = 0$ gives $E = E_g$.

As shown in Extended Data Fig. 3, the ratio $(E - E_g)/\sigma^2$ also remains nearly unchanged for different lattice sizes and symmetry sectors. This empirical conclusion is adopted to estimate the ratio in the large lattice from smaller ones so as to reduce the error and the time cost.

**Lanczos step.** The Lanczos step is a popular method in VMC for improving the variational accuracy[67]. It is also used in NQS[26,38].

The key idea of a Lanczos step is to construct new states $|\psi_p\rangle$ orthogonal to the well-trained variational wavefunction $|\psi_0\rangle$ and to minimize the energy of the new state formed by a linear combination of $|\psi_0\rangle$ and $|\psi_p\rangle$. The new energy is then guaranteed to be lower than the initial energy.

Only one Lanczos step is applied in this work, so we have one state $|\psi_1\rangle$ satisfying $\langle\psi_0|\psi_1\rangle = 0$ given by

$$|\psi_1\rangle = \frac{\mathcal{H} - E_0}{\sigma}\,|\psi_0\rangle, \tag{35}$$

where $E_0 = \langle\psi_0|\mathcal{H}|\psi_0\rangle$ and $\sigma^2 = \langle\psi_0|\mathcal{H}^2|\psi_0\rangle - E_0^2$. The linear combination of $|\psi_0\rangle$ and $|\psi_1\rangle$ can be written as

$$|\psi_\alpha\rangle = |\psi_0\rangle + \alpha\,|\psi_1\rangle, \tag{36}$$

whose energy is

$$E_\alpha = E_0 + \frac{\langle\psi_\alpha|(\mathcal{H} - E_0)|\psi_\alpha\rangle}{\langle\psi_\alpha|\psi_\alpha\rangle} = E_0 + \sigma\frac{\alpha^2\mu_3 + 2\alpha}{\alpha^2 + 1}, \tag{37}$$

where

$$\mu_n = \frac{\langle\psi_0|(\mathcal{H} - E_0)^n|\psi_0\rangle}{\sigma^n}. \tag{38}$$

The minimal energy is achieved at

$$\alpha_* = \frac{\mu_3 - \sqrt{\mu_3^2 + 4}}{2}, \tag{39}$$

and the lowest energy is

$$E_{\alpha_*} = E_0 + \sigma\frac{\alpha_*^2\mu_3 + 2\alpha_*}{\alpha_*^2 + 1} = E_0 + \sigma\alpha_*. \tag{40}$$

**Initial guess of $\alpha$.** A direct way to compute $\mu_n$ is by measuring suitable quantities as expectation values of the initial state $|\psi_0\rangle$. However, the measurement becomes more accurate if it is performed with a state $|\psi_{\alpha_0}\rangle$ closer to the ground state[67].

In this paper, we estimate the suitable $\alpha_0$ to obtain a $|\psi_{\alpha_0}\rangle$ closer to the true ground state compared to $|\psi_0\rangle$. Then, from equation (37), one can compute $\mu_3$ as

$$\mu_3 = \frac{\left(\alpha_0^2 + 1\right)\left(E_{\alpha_0} - E_0\right)/\sigma - 2\alpha_0}{\alpha_0^2}, \tag{41}$$

where $E_{\alpha_0}$ can be measured by Monte Carlo sampling. The optimal $\alpha_*$ can be derived from $\mu_3$ by equation (39), and the lowest energy is then given by equation (40).

**Energy variance.** To compute the energy variance of $|\psi_\alpha\rangle$, we start with an intermediate quantity

$$v_\alpha = \frac{\langle\psi_\alpha|(\mathcal{H} - E_0)^2|\psi_\alpha\rangle}{\sigma^2\langle\psi_\alpha|\psi_\alpha\rangle} = \frac{\alpha^2\mu_4 + 2\alpha\mu_3 + 1}{\alpha^2 + 1}. \tag{42}$$

Like the previous case, one can measure $v_{\alpha_0}$ by Monte Carlo sampling and determine $\mu_4$ as

$$\mu_4 = \frac{\left(\alpha_0^2 + 1\right)v_{\alpha_0} - 2\alpha_0\mu_3 - 1}{\alpha_0^2}. \tag{43}$$

Then $v_{\alpha_*}$ can be computed given $\mu_3$ and $\mu_4$, which gives the required energy variance as

$$\sigma_{\alpha_*}^2 = \frac{\langle\psi_{\alpha_*}|\left(\mathcal{H} - E_{\alpha_*}\right)^2|\psi_{\alpha_*}\rangle}{\langle\psi_{\alpha_*}|\psi_{\alpha_*}\rangle} = \sigma^2 v_{\alpha_*} - \left(E_0 - E_{\alpha_*}\right)^2. \tag{44}$$

## Data availability

This research does not rely on any external datasets. The data shown in Figs. 2 and 3 and the obtained neural network weights are available via Zenodo at https://zenodo.org/doi/10.5281/zenodo.7657551 (ref. 68).

## Code availability

We provide the code needed to reproduce our main results via Zenodo at https://zenodo.org/doi/10.5281/zenodo.7657551 (ref.68) and via GitHub at https://github.com/ChenAo-Phys/MinSR.

## Acknowledgements

We gratefully acknowledge M. Schmitt for help in improving the manuscript. We also thank T. Neupert, C. Roth, M. Bukov, F. Vicentini, W.-Y. Liu and X. Liang for fruitful discussions. This project has received funding from the European Research Council under the European Union's Horizon 2020 research and innovation programme (Grant Agreement No. 853443). This work was supported by the German Research Foundation (Project No. 492547816; TRR 360). We also acknowledge the Gauss Centre for Supercomputing e.V. (www.gauss-centre.eu) for funding this project by providing computing time through the John von Neumann Institute for Computing on the GCS Supercomputer JUWELS at Jülich Supercomputing Centre. We gratefully acknowledge the scientific support and high-performance computing resources provided by the Erlangen National High Performance Computing Center (NHR) of the Friedrich-Alexander-Universität Erlangen-Nürnberg (NHR Project No. nqsQuMat). NHR funding is provided by federal and Bavarian state authorities. NHR@FAU hardware is partially funded by the German Research Foundation (Grant No. 440719683).

## Author contributions

A.C. proposed the method and performed numerical simulations. M.H. provided computing resources and supervised the simulations. Both authors took part in the design of numerical experiments, the writing of the manuscript and the critical review of the paper.

## Funding

Open access funding provided by Universität Augsburg.

## Competing interests

The authors declare no competing interests.

## Additional information

**Extended data** is available for this paper at https://doi.org/10.1038/s41567-024-02566-1.

**Supplementary information** The online version contains supplementary material available at https://doi.org/10.1038/s41567-024-02566-1.

**Correspondence and requests for materials** should be addressed to Ao Chen or Markus Heyl.

**Peer review information** *Nature Physics* thanks Guglielmo Mazzola and the other, anonymous, reviewer(s) for their contribution to the peer review of this work.

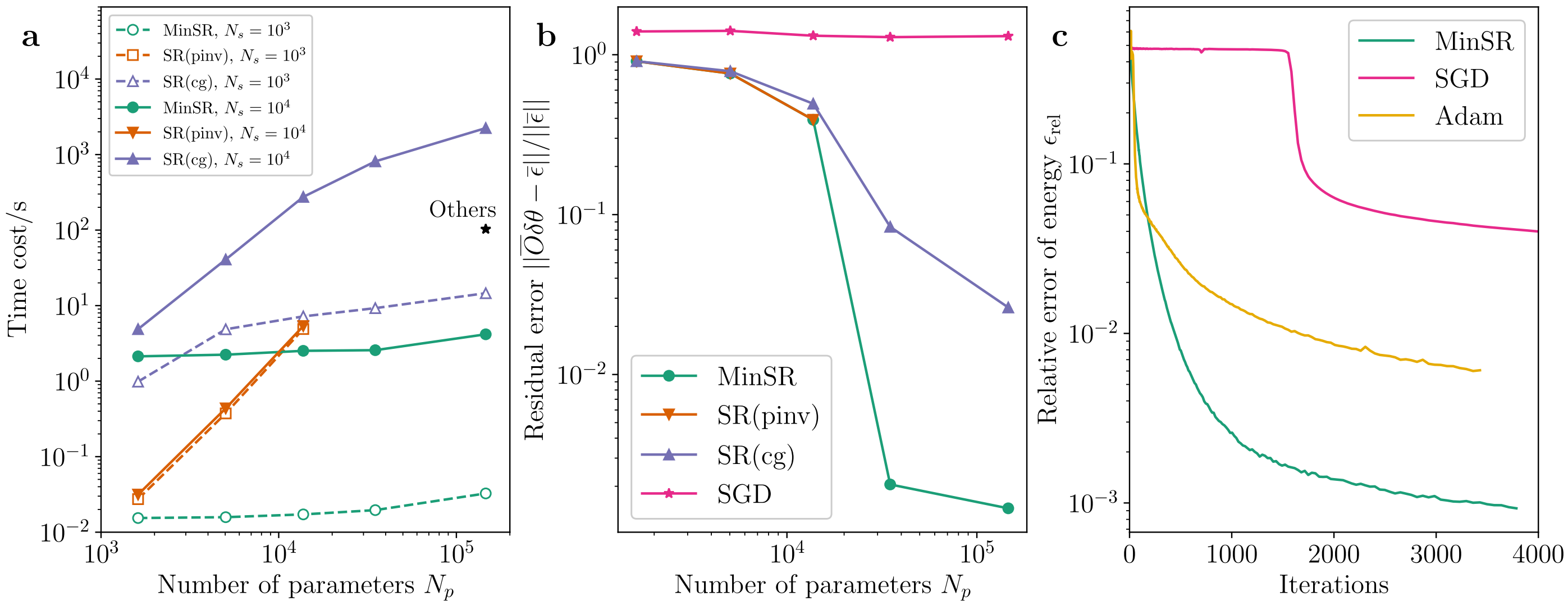

**Extended Data Fig. 1 | Performance evaluation of various optimization methods on the 10 × 10 square Heisenberg $J_1$-$J_2$ model. a,** Time cost of solving the equation $\overline{O}\delta\theta = \overline{\epsilon}$ for the different optimization methods for different numbers of Monte-Carlo samples $N_s$ and variational parameters $N_p$, measured at $J_2 = 0$ on the ResNet1 architecture and an A100 80G GPU. The time cost of other contributions in a training step on 4 parallel A100 GPUs is presented as the black star, showing that the time cost of SR becomes the bottleneck for training deep networks if MinSR is not employed. **b,** Relative residual error $||\overline{O}\delta\theta - \overline{\epsilon}||/||\overline{\epsilon}||$ for $N_s = 10^4$ at $J_2 = 0$ on ResNet1. **c,** Training curvea of ResNet2 at $J_2/J_1 = 0.5$ with $N_s = 10^4$ and $N_p \approx 10^6$ comparing different optimizers.

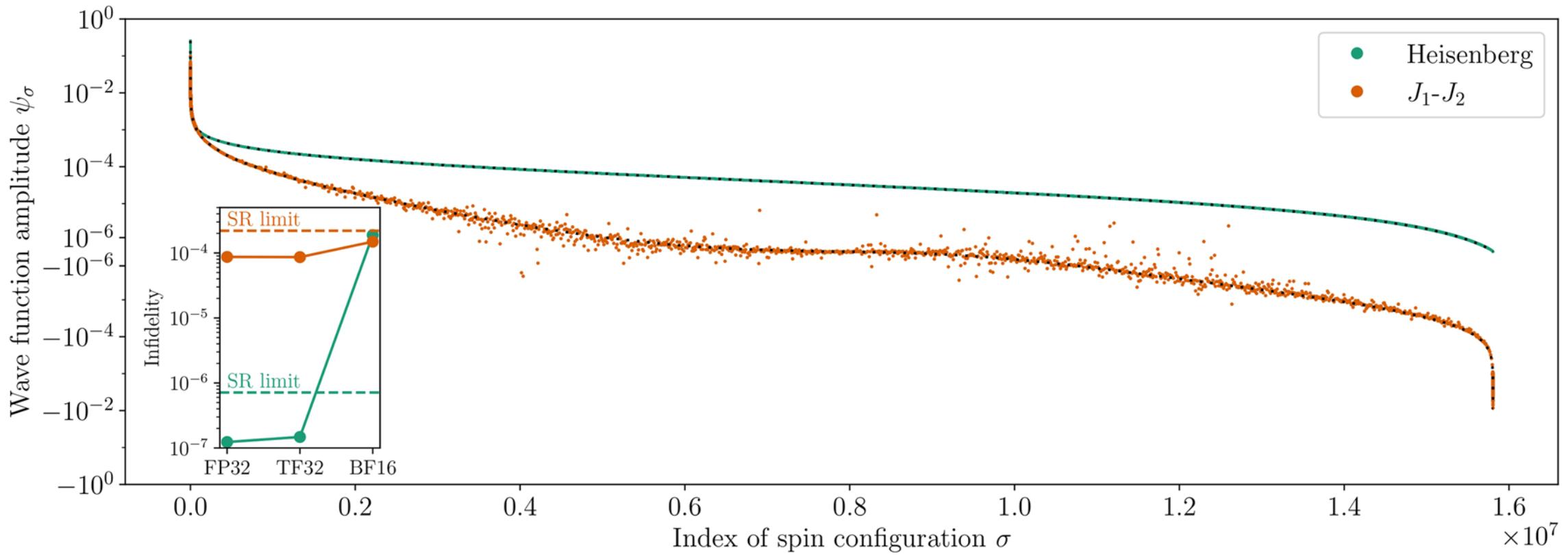

**Extended Data Fig. 2 | NQS wave function amplitudes for the Heisenberg and $J_1$-$J_2$ models on a 6 × 6 square lattice obtained by ResNet1 with 64 layers and 146320 parameters by means of MinSR.** The ED wave function amplitudes obtained by the `lattice-symmetries` package [64] are shown as the black dotted lines, and the spin configurations are sorted according to the descending order of ED amplitudes. All wave function amplitudes are shown for the Heisenberg model, while for the $J_1$-$J_2$ model only one point is plotted among 10000 successive points. In the inset, we show the infidelity with different numerical precision. The infidelity of a ResNet1 with 13750 parameters which approaches the size limit of SR (pinv) is also presented for comparison. This shows that the deep NQS trained by MinSR can approach TF32 precision in the non-frustrated case and BF16 precision in the frustrated case on the 6 × 6 lattice with a Hilbert space dimension of 15804956 after applying symmetry, while the shallow NQS trained by traditional SR cannot. In our further tests, the shallow network trained by traditional SR can only approach such precision on the 4 × 6 lattice with a much smaller Hilbert space dimension of 15578.

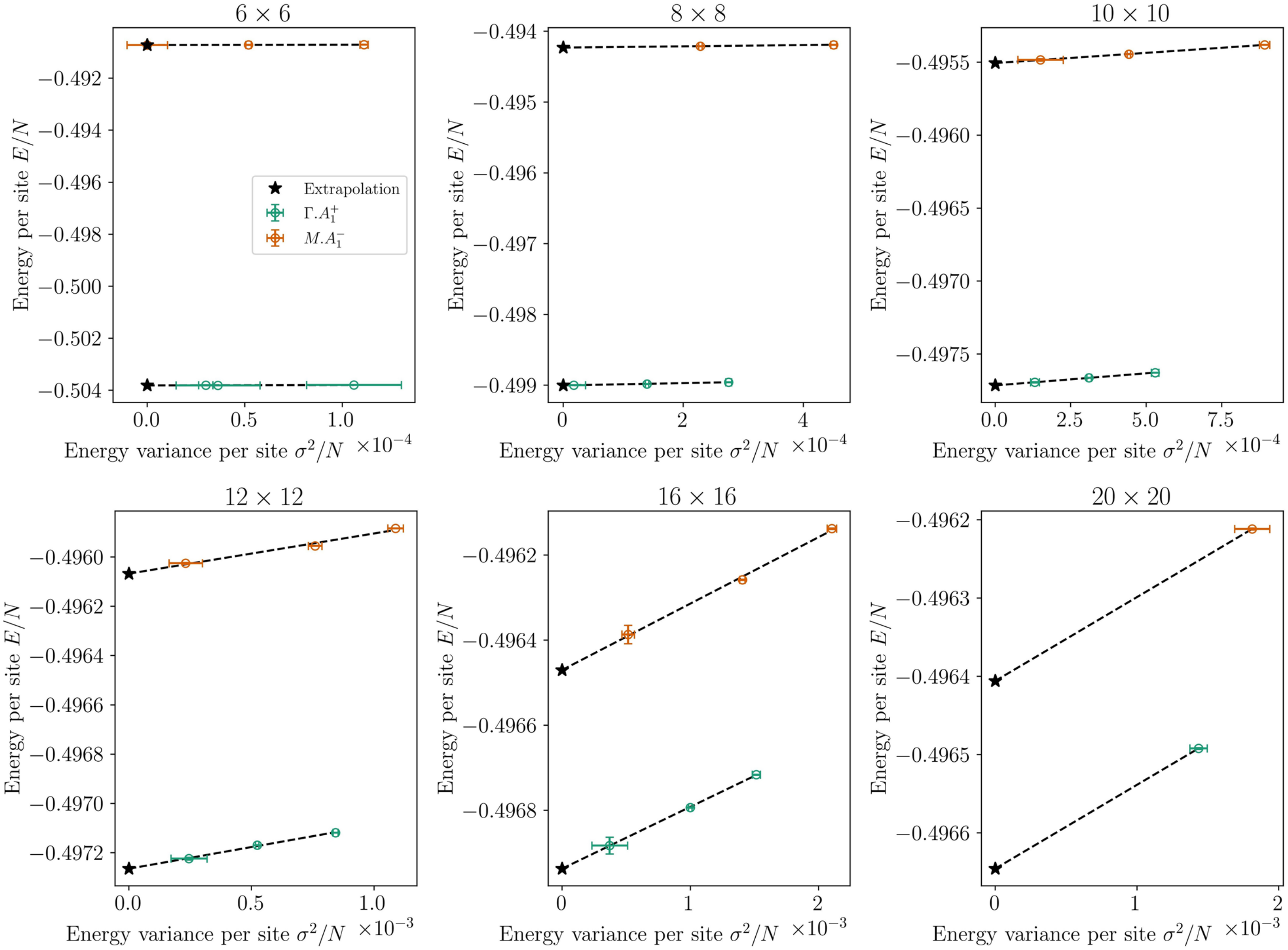

**Extended Data Fig. 3 | Zero-variance extrapolation for the square $J_1$-$J_2$ model at $J_2/J_1$ = 0.5 with different lattice sizes.** The three data points in the same fitting are obtained by two different sizes of real-valued ResNet2 with 34944 and 139008 parameters and a Lanczos step on the larger one. The ground state sector is $S = 0$ and $k = \Gamma = (0, 0)$, and the excited state sector is $S = 1$ and $p = M = (\pi, \pi)$. The error bars show the standard deviations. On the largest 20 × 20 lattice, the estimation of the slope is inaccurate in direct linear fitting due to the uncertainty of data points. Consequently, we utilize an empirical assumption that the slope remains nearly unchanged for different system sizes and symmetry sectors. Excluding too small lattices in which the estimation of slopes is inaccurate due to too close data points, the tendency of unchanged slopes is obvious for the slopes on $L$ = 10, 12, 16, respectively 0.17(2), 0.18(4), 0.15(4) for $S$ = 0, and 0.14(3), 0.16(2), 0.16(2) for $S$ = 1. The unchanged slope is also observed in existing literature [8, 16, 37]. Consequently, we employ the average of the aforementioned slopes as the slope on $L$ = 20 to mitigate the error.

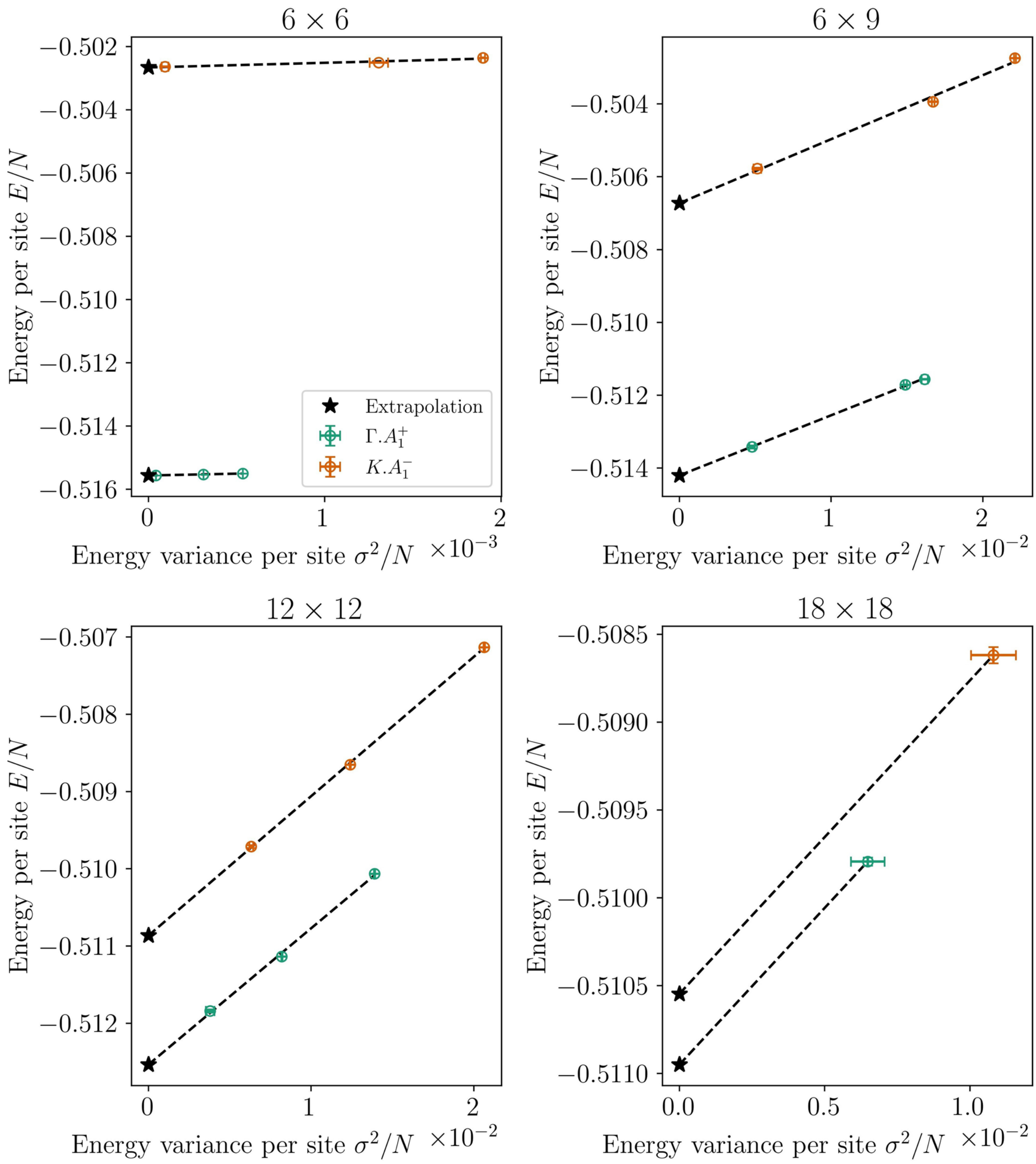

**Extended Data Fig. 4 | Zero-variance extrapolation for the triangular $J_1$-$J_2$ model at $J_2/J_1 = 0.125$ with different lattice sizes.** The three data points in the same fitting are obtained by two sizes of complex-valued ResNet2 with 34944 and 139008 parameters and a Lanczos step on the larger one. The ground state sector is $S = 0$, $k = \Gamma = (0, 0)$, and the excited state sector is $S = 1$, $k = K = (4\pi/3, 0)$. The error bars show the standard deviations. Similar to the square lattice case, the slope on the 18 × 18 lattice is approximated by the slope on the 12 × 12 lattice to mitigate the error. The data on the 18 × 18 lattice is generated by networks with 139008 parameters and a Lanczos step.

**Extended Data Table 1 | Variational ground state energies in the 10×10 square $J_1$-$J_2$ model with PBC at $J_2/J_1$=0.5**

| Wave function | Parameters | Reference | $E/N$ | $\sigma^2/N$ |
|---|---|---|---|---|
| **Tensor Network** | | | | |
| DMRG | 8192 SU(2) or 32000 U(1) states | [9] | -0.495530 | |
| PEPS + CNN | Bond dimension 4 in PEPS + 3531 real in CNN | [61] | -0.495502 | |
| **GWF** | | | | |
| Determinant | 5 real | [8] | -0.49521(1) | $\approx 7 \times 10^{-3}$ |
| Determinant + 2LS | 5 real | [8] | -0.49755(1) | $\approx 2 \times 10^{-3}$ |
| Determinant + RBM | 810 complex | [62] | -0.49575(3) | |
| PP + RBM | 10000 real in PP + 1616 complex in RBM | [24] | -0.497629(1) | $4.9(2) \times 10^{-4}$ |
| **Pure NQS** | | | | |
| CNN | 3838 complex | [31] | -0.49516(1) | |
| RNN | Unknown | [34,63] | -0.49560(4) | $2.9 \times 10^{-3}$ |
| RBM | 30000 complex | [38] | -0.49580(2) | |
| RBM + LS | 30000 complex | [38] | -0.4968(4) | |
| CNN + LS | 106529 real | [37] | -0.497468 | $\approx 1.4 \times 10^{-3}$ |
| GCNN | ≈ 75000 real | [26] | -0.497437(7) | |
| ViT(MinSR) | 267720 real | [64] | -0.497634(1) | |
| ResNet(MinSR) | 1071488 real | This work | -0.4976921(4) | $1.80(4) \times 10^{-4}$ |

As a reference, the ground state energy estimated by zero-variance extrapolation in this work is −0.497715(9).